\documentclass[aps,prc,showpacs,amsmath,amssymb,floatfix]{revtex4}
\usepackage{graphicx}
\usepackage{epsfig}
\usepackage{amsmath}
\begin{document}

\title{``Modified" BCS theory at finite temperature and its flaws}

\author{V.~Yu.~Ponomarev$^{1,2}$ and A.~I.~Vdovin$^{1}$}
\affiliation{
$^1$Bogoliubov Laboratory of Theoretical Physics,
Joint Institute for Nuclear Research,
141980 Dubna, Russia \\
$^2$Institut f\"ur Kernphysik, Technische Universit\"at Darmstadt,
D--64289  Darmstadt, Germany}
\date{\today }

\begin{abstract}
Drawbacks of a recently suggested modified BCS theory at finite
temperature in [Phys. Rev. C {\bf 67}, 014304 (2003)]
are discussed. We conclude that the thermal behavior of the pairing
correlations is not under control within this model.
\end{abstract}


\maketitle

A modified BCS (MBCS) model has been suggested recently~\cite{da}
to improve the BCS theory at finite temperature for nuclei.
It has been claimed that the MBCS ``{\it washes out the
sharp superfluid-normal phase transition}'' and ``{\it the fluctuations
of particle number also become more suppressed especially at high
temperature}''. In this Comment we argue that the corresponding
equations cannot be treated as an extension of the modified
BCS approach to finite temperature. Consequently, the ``physical effects"
reported in this article are only the results of incorrect physical
treatment and inconsistencies.

The MBCS equations at $T\neq0$ are obtained applying a 
temperature-dependent 
unitary transformation to the Bogoliubov quasiparticles
$\{\alpha^{\dagger}_{jm},\ \alpha_{jm}\}$, thus transforming them into 
the new
bar quasiparticles $\{\bar{\alpha}^{\dagger}_{jm},\ \bar{\alpha}_{jm}\}$:
\begin{equation}
\bar{\alpha}^{\dag}_{jm} = \sqrt{1-n_j} \alpha^{\dag}_{jm} +
(-1)^{j-m} \sqrt{\vphantom{1}n_j} \alpha_{j -m},
\label{e1}
\end{equation}
where $n_j$ are the
thermal Fermi-Dirac occupation numbers for the Bogoliubov
quasiparticles. The authors claim: ``{\it The
MBCS equations $\ldots$ has exactly the same form as the
standard BCS equations}
$
 (\Delta_{T=0} = G \sum_j \Omega_j u_j v_j)
$
{\it where 
the coefficients $u_j$ and $v_j$
are replaced with $\bar{u}_j$ and  $\bar{v}_j$''}:
\begin{eqnarray}
{\bar \Delta} (T) &=& G \sum_j \Omega_j {\bar u_j} {\bar v_j} =
G \sum_j \Omega_j \bigg[(1-2n_j)u_j v_j  \nonumber \\
& - &  \sqrt{n_j (1-n_j)} (u_j^2 -v_j^2) \bigg]~~.
\label{gap}
\end{eqnarray}

Although formally the conventional BCS ($T=0$) and MBCS equations look 
similar, there is one essential difference between them. As far as 
$u_j$ and $v_j$ coefficients are positively defined, the BCS gap
$\Delta = 0$ (breaking of the Cooper pairing) corresponds to the trivial
solution $\{u_j,~v_j\} = \{0(1),1(0)\}$ for all $j$ when all 
levels give zero contribution to the gap.
It is not difficult to show analytically that the MBCS gap equation 
(\ref{gap}) does not have such a trivial solution 
$\{{\bar u}_j,~{\bar v}_j\} = \{0(1),1(0)\}$. 
Indeed, from Eq.~(38) in Ref.~\cite{da}:
\begin{equation}
{\bar u}_j = u_j \sqrt{1-n_j} + v_j \sqrt{\vphantom{1}n_j},~~~
{\bar v}_j = v_j \sqrt{1-n_j} - u_j \sqrt{\vphantom{1}n_j}
\label{uv}
\end{equation}
the trivial solution $\{{\bar u}_j,~{\bar v}_j\} = \{0(1),1(0)\}$
corresponds to:
\\[1mm]
\begin{tabular}{lll}
\hspace*{46mm}
$u_j = \sqrt{1-n_j}$; & ~~~$v_j = \sqrt{\vphantom{1}n_j}$ &~~~-- particles \\
\hspace*{46mm}
$u_j = -\sqrt{\vphantom{1}n_j}$; & ~~~$v_j = \sqrt{1-n_j}$& ~~~-- holes
\end{tabular}
\\[1mm]
and contradicts the positive definition of $u_j$.

One may notice from Eq.~(\ref{uv}) that ${\bar v_j}$ coefficients 
become negative above a certain temperature for particle 
levels since $v_j << u_j$ for them.
Then the MBCS pairing gap $\bar{\Delta}$ receives positive contribution 
from hole levels and negative contribution from particle levels.
This leads to quite a strange thermal behavior of $\bar \Delta (T)$.

On page 5 of Ref.~\cite{da} we find a discussion of the MBCS critical
temperature ${\bar T}_c$ defined as ${\bar \Delta}({\bar T}_c) =0$.
According to the authors this temperature is reached 
when the ``new'' (second) term in
Eq.~(\ref{gap}) is equal to the conventional one.
Numeric calculations (see below) show that two terms in Eq.~(\ref{gap})
compensate each other around the critical temperature of the conventional
BCS $T_c \approx 0.57 \cdot \Delta_{T=0}$ for particle levels and they
never do for hole levels because the second term in Eq.~(\ref{gap})
is always negative for holes. 

As clear from the above paragraphs, 
${\bar \Delta}$ vanishes at ${\bar T}_c$  only
when a negative contribution from particles and
positive contribution from holes cancel each other (notice the difference
with the conventional BCS $\Delta =0$).
However if this happens, at higher $T$ the balance appears to be broken
and ${\bar \Delta}$ becomes finite again.

\begin{figure}
\epsfig{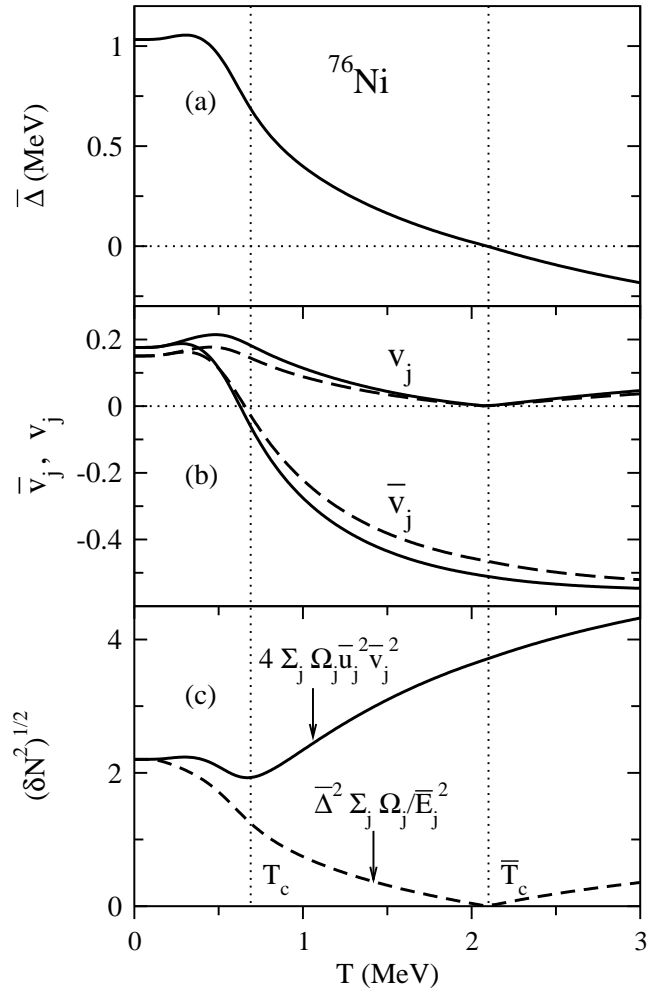}
\caption{\label{f1}
(a) the MBCS pairing gap;
(b) $\bar{v}_j$ and $v_j$ coefficients for particle
levels near the Fermi surface: 2d$_{5/2}$ - solid lines,
3s$_{1/2}$ - dashed lines;
(c) two functions for particle-number fluctuations $\sqrt{\delta N_i^2}$
in Eq.~(\ref{e3}).
Critical temperates
$T_c$ and ${\bar T}_c$  are indicated by vertical lines.
See text for details.
}
\end{figure}

To be more specific,
we plot in Fig.~\ref{f1}a the MBCS pairing gap as function of temperature
in $^{76}$Ni
\footnote{Our code reproduces excellently all the results in Ref.~\cite{da}
for all isotopes. We present them for $^{76}$Ni only (from
the resonant continuum MBCS calculations). The ones in $^{68-76}$Ni
and $^{80-82}$Ni are qualitatively very similar and lead to the same
conclusions.}.
One notices that it reaches zero at ${\bar T}_c \approx 2.1$~MeV
and continues to decrease. The authors of Ref.~\cite{da} define an
unphysical negative
gap of the MBCS as the results ``{\it no longer reliable}''.
Since the authors never discuss any physical limitations of their model,
it is not clear upon what basis this conclusion is made.
In Fig.~\ref{f1}b we present the ${\bar v}_j$ and $v_j$ coefficients for 
two particle (2d$_{5/2}$, 3s$_{1/2}$)
levels near the Fermi surface. They become negative around $T_c$.
However, one may notice that nothing special occurs
at ${\bar T}_c$, thus confirming the
``unreliability" of the method above the ``new" critical temperature. 

The problem with a negative gap can be removed by increasing the
role of hole levels. It may be achieved, e.g., by including
the levels from the shell N=0-28
in a single-particle spectrum in addition
to the levels near the Fermi surface accounted for in Ref.~\cite{da}
(with a proper renormalization of the pairing strength).
Then, another MBCS crack appears: the gap starts
to continously increase above a certain temperature remaining always 
positive. Similar effect in $^{84}$Ni is shown in Figs. 2h and 3 of
Ref.~\cite{da}.
But we find it impossible for the MBCS gap to reach zero and remain zero at
higher temperature, thus, indicating the superfluid-normal phase transition.

In view of all the above, the author's conclusions about superfluid-normal 
phase transition within the MBCS makes no sense from our point of view.

The other main goal of the MBCS is to reduce the particle number fluctuations
in the ground state. The authors show in Fig.~5
(solid line) of Ref.~\cite{da} that this quantity
\begin{equation}
\delta \bar{N}^2 = 4 \sum_j \Omega_j \bar{u}_j^2 \bar{v}_j^2  =
\bar{\Delta}^2 \sum_j \Omega_j / \bar{E}_j^2.
\label{e3}
\end{equation}
is reducing with the temperature and apparently vanishes.
The $\delta \bar{N}^2$  has been calculated from the r.h.s. part of
Eq.~(\ref{e3}) (or Eq.(41) in Ref.~\cite{da}). It remains unclear how
transformation in Eq.~(\ref{e3}) has been obtained within the MBCS
at $T \ne 0$.
We have calculated both the left and right functions.
Numeric results presented in Fig.~\ref{f1}c reveal their rather different
behavior.
It means that the conclusion on suppression of the particle number
fluctuations with a temperature within the MBCS in Ref.~\cite{da}
is based on calculations from the equation which is not valid.

A legitimate question is what is wrong with the MBCS.
In Sec. II.D and Appendix A of Ref.~\cite{da} the
well-known  statistical approach  in a many-body theory at finite
temperature is sketched.  However, deriving the MBCS equations
the authors of Ref.~\cite{da} do not use this method.
Instead, they formally follow the
scheme formulated by them for zero temperature. Namely,
they find a minimum of the expectation value  of the BCS
Hamiltonian over the temperature dependent vacuum state $|\bar{0}
\rangle$. 
They imply for reasons unclear that selecting
$n_j(T)$ in the form of the quasiparticle thermal occupation numbers
is equivalent to averaging over the grand canonical ensemble
(as one reads on page 5 of Ref.~\cite{da}).

We find it difficult to understand:
the authors are aware that a quantum mechanical ground state,
for which expectation value of the Hamiltonian equals the 
statistical average over the grand canonical ensemble,
``{\it does not exist in the physical space
spanned by eigen vectors}'' of the pairing Hamiltonian and why it is so
(see footnote on page 3).
Nevertheless, the basic starting point of the MBCS at $T\neq0$
is certainly to introduce such a vacuum state $|{\bar 0} \rangle$ in Eq.~(33).

This means that the MBCS cannot be treated as a theory with
the thermal behavior of the pairing correlations under control.
It is not surprising that the MBCS equations (Eqs.~(39-40) and (43-44)
in Ref.\cite{da}) differ from the conventional ones by extra terms
but their origin is only inconsistency in treatment of the physical
object.

To conclude, we doubt that the modified BCS in Ref.~\cite{da} with the
above discussed flaws is a step forward in the developing of the 
conventional BCS at finite temperature.

\end{document}